\documentstyle[12pt]{article}

\newcommand{\be}{\begin{equation}}   \newcommand{\ee}{\end{equation}}
\newcommand{\bear}{\begin{eqnarray}}
\newcommand{\eear}{\end{eqnarray}}
\newcommand{\ba}{\begin{array}}      \newcommand{\ea}{\end{array}}
\newcommand{\lae}{\begin{array}{c}\,\sim\vspace{-21pt}\\< \end{array}}

\begin{document}

\pagestyle{empty}
\begin{titlepage}
\def\thepage {}        

\title{\bf Electroweak Symmetry Breaking\\ [2mm] via Top Condensation
Seesaw
\\ [1cm]}

\author{\bf Bogdan A.~Dobrescu$^1$ and  Christopher T.~Hill$^{1,2}$ \\
\\
{\small {\it $^1$Fermi National Accelerator Laboratory}}\\
{\small {\it P.O. Box 500, Batavia, Illinois, 60510, USA \thanks{e-mail
  addresses: bdob@fnal.gov, hill@fnal.gov} }}\\
\\
{\small {\it $^2$ The University of Chicago }}\\
{\small {\it Chicago, Illinois, 60637, USA }}\\ }

\date{ }

\maketitle

\vspace*{-12cm}
\noindent
\makebox[10.7cm][l]{hep-ph/9712319} FERMILAB-PUB-97/409-T \\ [1mm]
\makebox[10.7cm][l]{December 19, 1997} EFI--97--55 \\ [1mm]
\makebox[10.7cm][l]{Revised June 26, 1998} \\

 \vspace*{13.1cm}

\baselineskip=18pt

\begin{abstract}

   {\normalsize
We propose a new mechanism in which all of the
electroweak symmetry breaking
is driven by a natural top quark condensate, produced by topcolor
interactions at a multi-TeV scale. The scheme 
allows the observed top quark mass, and acceptable $T$ and $S$ parameters,  
by invoking a seesaw mechanism involving mixing of the 
top with an additional isosinglet quark. 

}

\end{abstract}

\vfill
\end{titlepage}

\baselineskip=18pt
\pagestyle{plain}
\setcounter{page}{1}


The proximity of the measured top quark mass, $m_t$, to the electroweak
scale hints that electroweak symmetry breaking  (EWSB)
has its origin in dynamics associated with the top quark.
An explicit realization of this idea is the top condensation mechanism 
\cite{bhl}, in which the top---anti-top quark pair acquires a vacuum 
expectation value, much like the chiral condensate of QCD or the 
electron pairing condensate of BCS superconductivity.  The
$\langle \bar{t} t \rangle$ condensate has
the correct quantum numbers for EWSB, and
it connects the electroweak scale directly to $m_t$.
However, the original schemes of this type \cite{bhl, cond}
are problematic: (1) they require the scale of new physics to be very large, 
of order the GUT scale, $\sim 10^{15}$ GeV; (2) they require an unnatural  
cancellation of
large quadratic mass scales, (3) they lead in the standard model to the RG 
fixed point prediction $m_{t} \sim 230$ GeV which is too high 
\cite{CTH}  (note though that in the minimal
supersymmetric standard model, the fixed point
prediction is reasonable \cite{mssm}).

Subsequently, a specific gauge dynamics, ``topcolor", was proposed to
generate a top condensate at the TeV scale \cite{topcolor}. This model
views the heaviness of the top quark as a dynamical phenomenon which is
essentially independent of EWSB, 
provides a specific model of the new dynamics, and attempts
to place the scale of the dynamics at $\sim 1$ TeV to avoid fine--tuning
issues. It produces associated pseudo--Nambu-Goldstone
bosons with decay constant $f_{t\pi}$. Using the large-$N$, fermion loop  
approximation one estimates:
\be
f_{t\pi}^2 \approx \frac{3}{4 \pi^2} m_{t}^2 
\ln \left(\frac{M}{m_t}\right) ~.
\ee
With $M \sim 1$ TeV we obtain $f_{t\pi}\approx 64$ GeV (EWSB, 
would require $f_{t\pi} = 246$ GeV).
 Hence, topcolor has
always been combined
with additional dynamics, such as technicolor  \cite{tc2}.  
Indeed, this provides a solution to the serious problem 
of the large top quark mass
within the context of technicolor models, and interesting
Topcolor-Assisted Technicolor models (or TC$^2$) have been constructed 
\cite{ntc2}. These models, though potentially viable, are somewhat
cumbersome and implications of limits 
on custodial symmetry violation \cite{cdt}, and 
other phenomenological constraints \cite{fcnc,ct} require dynamical 
assumptions which are difficult to analyze. 

In the present letter we propose a new mechanism, in which the EWSB
occurs via the condensation of the top quark 
in the presence of an extra vectorlike, weak-isosinglet quark. 
The mass scale of the condensate is large, of order 0.6 TeV
corresponding to the electroweak scale $v \approx 246$ GeV.
The vectorlike isosinglet then naturally admits
a seesaw mechanism, yielding the physical top quark mass, which is then
adjusted to the experimental value.  The choice of a natural multi-TeV 
scale for the topcolor dynamics then determines the mass of the 
vectorlike quark.

There are several attractive features of this mechanism:
(i) The model is relatively specific, the left--handed top quark being 
unambiguously  identified as the electroweak--gauged condensate fermion, 
and the scheme uniquely specifies topcolor, together with some new $U(1)$
interactions, as the primary new strong interaction; (ii) the scheme is 
economical, requiring no additional weak--isodoublets, and therefore
easily satisfies the constraints
upon the $S$ parameter, using estimates made in the large--$N$ approximation; 
(iii) the  constraint
on custodial symmetry violation, i.e., the value of the 
$\Delta \rho_* \equiv \alpha T$ parameter, is easily satisfied,
being principally the usual $m_t$ contribution suppressed by the squared
ratio of the mixings between the top and the vectorlike quark.
We mention that we were led to these schemes by considering 
fermionic bound states in the context of strong dynamical models. We will
not elaborate this aspect presently, but note that the vectorlike  
quark in our model may potentially arise from within the dynamics without 
being introduced ad-hoc (an example of this type is given in 
ref.~\cite{suzuki}).

Consider the embedding of the standard model into a ``topcolor'' scheme:
$SU(3)_{1} \times SU(3)_{2} \times SU(2)_{W} \times U(1)_{1} 
\times U(1)_{2} \times U(1)_{B-L}$ gauge group where the 
$SU(3)_{1} \times  U(1)_{1}$ ($SU(3)_2 \times  U(1)_2$)
acts only on the third (first and second) generation quarks and leptons.
The $U(1)_{B-L}$ charges are $x < 0$ for $t_R$, 1/3 for the other quarks, and 
$-1$ for leptons (including right-handed neutrinos).
In addition to the observed quarks and leptons
we include a new fermion, $\chi$, whose left- and right-handed components 
transform identically to $t_R$, except 
that the $U(1)_{B-L}$ charge of $\chi_L$ is $x$, such that the anomalies 
cancel.
At a multi-TeV scale the gauge symmetries are broken
as follows: (1) $SU(3)_{1} \times SU(3)_{2}\rightarrow SU(3)_{QCD}$,
leaving a degenerate octet of massive ``colorons'';
(2) $U(1)_{1}  \times U(1)_{2} \rightarrow 
U(1)_{Y}$, and (3) $ U(1)_{B-L}$ is broken, 
leaving two heavy gauge bosons, $Z_1$ and $Z_{B-L}$; (4) $SU(2)_W$ is unbroken.

We emphasize that this specific choice of the
gauge groups and particle content is considered presently
for the purpose of illustration. To achieve the
desired physics in this scheme will require some fine-tuning
of the gauge coupling constants to a few percent of critical values.  
We believe that this fine-tuning can be alleviated
in more general schemes, and in fact, we expect to generalize the
mechanism when it is extended to give the light quark and lepton masses.
Moreover, we will work in the fermion bubble approximation to leading order
in $N_c$, the number of colors.  This is a crude approximation
to the dynamics, so the specific estimates made here are 
expected to be reliable to roughly only a factor of two or so. 

Integrating out the colorons yields an effective Lagrangian
for the third generation fields:
\be
{\cal L}_{\rm eff}^c =
- \frac{4\pi \kappa}{M^2}\left(
\overline{q} \gamma_\mu \frac{\lambda^{A}}{2} q 
+  \overline{\chi} \gamma^\mu \frac{\lambda^{A}}{2} \chi 
\right)^2 ~,
\label{ops1}
\ee
where $q=(t,b)$, the coloron mass is $M$, 
$\kappa = (g^2_s \cot^2\theta)/8\pi$
where $g_s$ is the QCD coupling, and
$\theta$ is the usual $SU(3)_{1} \times SU(3)_{2}$ mixing angle 
\cite{topcolor,tc2} (the first and second generations
feel effects that are proportional to $\tan^2\!\theta$ and mixing with
the third generation that are of order $1$); we assume $\cot\theta \gg 1$.
For critical $\kappa $ the effects of $ {\cal L}_{\rm eff}^c $ 
alone would produce an $ SU(3) $ symmetric condensate with 
$\langle\overline{t} t\rangle = \langle\overline{b} b\rangle = 
\langle\overline{\chi} \chi \rangle$, 
and an $SU(3)$ octet of Nambu--Goldstone bosons (NGB's).

Integrating out the two massive $U(1)$ bosons
yields the effective Lagrangians:
\bear
{\cal L}_{\rm eff}^1 & = &
- \frac{4\pi \kappa_1}{M_1^2}\left(
\frac{1}{3}\overline{q}_L \gamma_\mu  q_L
+ \frac{4}{3}\overline{t}_R \gamma_\mu t_R
- \frac{2}{3}\overline{b}_R \gamma_\mu b_R
+  \frac{4}{3} \overline{\chi} \gamma^\mu \chi 
\right)^2 ~,
\label{ops2}
\eear
and
\bear
{\cal L}_{\rm eff}^{B-L} & = &
- \frac{4\pi\kappa_{B-L} }{M_{B-L}^2}\left(
\frac{1}{3}\overline{q}_L \gamma_\mu  q_L
+ x \overline{t}_R \gamma_\mu t_R
+ \frac{1}{3}\overline{b}_R \gamma_\mu b_R
+ x \overline{\chi}_L \gamma^\mu \chi_L 
+ \frac{1}{3} \overline{\chi}_R \gamma^\mu \chi_R
\right)^2 ~,
\label{ops3}
\eear
as well as four-fermion operators involving the third generation leptons.
Here $M_1$ and $M_{B-L}$ are the $Z_1$ and $Z_{B-L}$ masses, 
$\kappa_1 = g^{\prime 2} \cot^2\theta^\prime /(8 \pi)$, $g^\prime$ is 
the hypercharge gauge coupling, $\theta^\prime$ is the mixing angle between the
$U(1)_{1} \times U(1)_{2}$ gauge bosons, 
$\kappa_{B-L} = g_{B-L}^{2} /(8 \pi)$, $g_{B-L}$ is the 
$U(1)_{B-L}$ gauge coupling. 
We shall assume, for simplicity, that all the massive gauge bosons have 
a common mass $M$. These interactions are typically 
attractive and non-confining. In addition, electroweak preserving mass 
terms are allowed in the low energy theory:
\be
{\cal L}_{\rm mass} = - \mu_{\chi\chi} \overline{\chi}_L\chi_R
- \mu_{\chi t} \overline{\chi}_L t_R ~.
\label{lmass}
\ee
Note that the $\overline{\chi}_L\chi_R$ mass term is not $U(1)_{B-L}$ invariant,
but it is induced below the 
scale $M$ if the $\chi$ couples to the $U(1)_{B-L}$ breaking VEV.

In the presence of the full effects of ${\cal L}_{\rm eff}^1 $
and ${\cal L}_{\rm eff}^{B-L}$, the theory can produce a different
pattern of condensates, i.e., a different pattern of chiral symmetry breaking,
than in the case of the pure ${\cal L}_{\rm eff}^c $. 
To see this we note that the mass-gap equations for the model
take the generic form
\be
m_{AB} = z_{AB} m_{AB} F_{AB}\!\left(m_{\chi\chi}, m_{\chi t}, 
m_{t \chi}, m_{t t}, ...
\right) ~,
\label{gaps}
\ee
where the indices $A$ and $B$ stand for $\chi$, $t$ and $b$, $m_{AB}$ 
is the dynamical 
mass generated by the condensate $\langle \overline{A}_L B_R \rangle$, 
and the functions $F_{AB}$ depend on all nine dynamical masses.
The coefficients $z$ are combinations of the 
$SU(3)_1\times U(1)_1 \times U(1)_{B-L}$
gauge couplings and charges:
\be
z_{AB} = \frac{2}{\pi} \left( \frac{4}{3}\kappa + Y_A Y_B \kappa_1 +  
X_A X_B \kappa_{B-L} \right) ~,
\ee
where $Y$ and $X$ are the $U(1)_1$ and $U(1)_{B-L}$ charges, respectively.
Our charge assignment gives the following inequalities:
\bear
& & z_{t\chi} > z_{t t}, \, z_{t b} > z_{\chi b} ~,
\nonumber \\ [1mm]
& & z_{\chi t} > z_{\chi\chi} > z_{t t} ~.
\label{inequalities}
\eear
A necessary condition for having a non-zero dynamical mass $m_{AB}$
is that at least one of the three $z_{A^\prime B}$ coefficients 
is above a critical value, $z_{\rm crit} = 1$, or that the corresponding
current mass $\mu_{AB}$ is non-zero. 
In the low energy effective theory 
this condition corresponds to the requirement of having a negative squared 
mass or a tadpole term for the composite scalar formed in the 
$\overline{A}_L B_R$ channel \cite{cdh}. 
We need the formation of the $\langle\overline{\chi}_L t_R\rangle$ and 
$\langle\overline{t}_L\chi_R\rangle$ condensates, and therefore we
require $z_{\chi t},z_{t \chi} > 1$. 
Choosing the three gauge couplings such that $z_{t b} < 1$
ensures that the $\overline{q}_L b_R$ and $\overline{\chi}_L b_R$ 
channels are sub-critical, so that $b_R$ does not participate in condensates.
Furthermore, if $z_{tt}$ is sub-critical, the 
$\bar{q}_L t_R$ condensate will be aligned with $\bar{q}_L \chi_R$.
In addition, we assume for convenience that
$z_{\chi\chi} < 1$. All the above conditions on the $z$ coefficients 
are satisfied provided
\be
{\rm max} \left(\frac{4\kappa_1}{3} + \frac{x\kappa_{B-L}}{4} , 
\; - \frac{\kappa_1}{6} + \frac{\kappa_{B-L}}{12} \right) < 
\frac{3\pi}{8} - \kappa < {\rm min} 
\left(\frac{\kappa_1}{3} + \frac{\kappa_{B-L}}{12}, 
\; \frac{4\kappa_1}{3} + \frac{3x^2}{4}\kappa_{B-L}\right)
\label{criticality}
\ee
In this case, the most general dynamical mass matrix (up to an $SU(2)_W$
transformation) is given in the weak
eigenstate basis by
\be
{\cal L} = - \left( \overline{t}_L \ , \ \overline{\chi}_L  \right) 
\left( \ba{rl}  m_{t t}  & m_{t \chi} \\ 
        m_{\chi t} & m_{\chi \chi} \ea \right) 
\left( \ba{c} t_R \\ \chi_R \ea \right) + {\rm h.c.} 
\label{massterm}
\ee
Therefore, we need to solve the subset of four coupled gap equations 
included in eq.~(\ref{gaps}), where the functions $F$ 
can be computed by keeping the weak 
eigenstates in the external lines, and the $\chi$ and $t$ mass eigenstates
running in the loop. Keeping only the quadratic and logarithmic divergences, 
with a physical cut-off given by the mass of the heavy gauge bosons, we 
obtain:
\bear
F_{\chi \chi} & = & 1 - \left(m_{t\chi}^2 + m_{\chi t}^2 + m_{\chi\chi}^2
+ \frac{m_{tt}m_{t \chi}m_{\chi t}}{m_{\chi\chi}} \right)
\frac{1}{M^2}\ln\left( \frac{M^2}{m^2_{\chi \chi}} \right)
+ \frac{\mu_{\chi\chi}}{z_{\chi\chi} m_{\chi\chi}}
   \nonumber \\ [3mm]
F_{\chi t} & = & 1 - \left( m_{tt}^2 + m_{\chi t}^2 + m_{\chi\chi}^2 
+ \frac{m_{tt}m_{t \chi}m_{\chi\chi}}{m_{\chi t}} \right)
\frac{1}{M^2}\ln\left( \frac{M^2}{m^2_{\chi \chi}} \right)
+ \frac{\mu_{\chi t}}{z_{\chi t} m_{\chi t}}
   \nonumber \\ [3mm]
F_{t \chi} & = & 1 - \left( m_{tt}^2 + m_{t\chi}^2 + m_{\chi\chi}^2
	+ \frac{m_{tt}m_{\chi t}m_{\chi\chi}}{m_{t \chi}} \right)
\frac{1}{M^2}\ln\left( \frac{M^2}{m^2_{\chi \chi}} \right) 
   \nonumber \\ [3mm]
F_{t t} & = & 1 - \left( m_{tt}^2 + m_{t\chi}^2 + m_{\chi t}^2
	+ \frac{m_{t\chi}m_{\chi t}m_{\chi\chi}}{m_{t t}} \right)
\frac{1}{M^2}\ln\left( \frac{M^2}{m^2_{\chi \chi}} \right)  ~.
\label{set}
\eear
In what follows we will be interested for simplicity in the case where 
the $\langle \overline{\chi}_L \chi_R \rangle$ condensate is significantly
larger than the other ones, such that the seesaw condition,
\be
 m_{\chi \chi}^2 \gg  m_{\chi t}^2 , m_{t \chi}^2 \gg 
\frac{m_{\chi \chi}^2}{m_{\chi t}^2} m_{tt}^2~,
\label{situation}
\ee
is satisfied. 
A nontrivial solution to the set of gap equations can be easily found in this 
case (we assume $\mu_{\chi t} \ll \mu_{\chi \chi}$): 
\bear
m_{\chi\chi} & \approx & \frac{\mu_{\chi\chi}}{z_{t\chi} - z_{\chi\chi}} > 0
   \nonumber \\ [3mm]
m_{\chi t, t\chi}^2 & \approx & \frac{M^2}{\ln 
\left(\frac{M^2}{m_{\chi\chi}^2}\right)}
\left(1 - \frac{1}{z_{\chi t, t\chi}} \right) 
- m_{\chi\chi}^2 > 0
   \nonumber \\ [3mm]
m_{tt} & \approx & - 
\frac{z_{tt}m_{\chi\chi}m_{\chi t}m_{t \chi}}
{(1 - z_{tt})M^2}
\ln \left(\frac{M^2}{m_{\chi\chi}^2}\right) < 0 ~.
\label{additional}
\eear
The seesaw condition (\ref{situation}) requires a partial cancellation of the 
terms in the above expression for $m_{\chi t, t\chi}^2$, and a constraint 
on $\mu_{\chi \chi}$,
\be
\frac{\mu_{\chi \chi}^2}{M^2} \ll
(1 - z_{tt})(z_{t\chi} - z_{\chi\chi})^2 ~.
\label{hierarchy}
\ee
It should be emphasized that 
it is possible to satisfy conditions (\ref{criticality}) and
 (\ref{situation}) without excessive fine-tuning. 
To see this, note that the $SU(3)_1$ interactions are likely to be
stronger than the $U(1)$'s, so that eq.~(\ref{criticality})
gives the criticality condition
\be
\kappa = \frac{3\pi}{8} \left[ 1 + 
O \left( \frac{\kappa_1}{\kappa}, \frac{\kappa_{B-L}}{12\kappa}  \right)\right] 
~.
\ee
According to the discussion in ref.~\cite{cdt}, the ratios $\kappa_1/\kappa$ 
and $\kappa_{B-L}/(12\kappa)$
can be interpreted as the amount of fine-tuning in $\kappa$ required by 
eq.~(\ref{criticality}). 
From this point of view, we consider values of order 0.1 -- 0.01 for these 
ratios to be acceptable.
We return to an example of the numerical values of coupling constants after 
obtaining some phenomenological constraints on the masses.

As a consequence of condition (\ref{situation}), the physical top mass is 
suppressed by a seesaw mechanism:
\be
m_t \approx \frac{m_{\chi t} m_{t \chi}}{  m_{\chi \chi}} \left[
1 + O\left(m_{\chi t, \, t\chi}^2/m_{\chi \chi}^2\right) \right]~.
\label{topmass}
\ee
The electroweak symmetry is broken by the $m_{t \chi}$
dynamical mass. Therefore, the electroweak scale is estimated to be
given by
\be
v^2 \approx \frac{3}{4 \pi^2} m_{t\chi}^2 
\ln \left(\frac{M}{m_{\chi\chi}}\right) ~.
\label{vev}
\ee
It is easy to verify that this estimate is correct to first order in
$(m_{t\chi}/m_{\chi\chi})^2$, by diagonalizing the mass matrix in 
eq.~(\ref{massterm}) and computing the one-loop leading contribution to the $W$
mass. This can also be seen in an effective Lagrangian analysis;
this term comes from a dynamically generated composite Higgs boson
with a very strong coupling $g \sim 4\pi/\sqrt{3\ln(M/m)}$
to $\bar{t}\chi$, with the usual VEV,
where eq.~(\ref{vev}) is just the usual $m = g v/2$ ``Goldberger-Treiman''
relation.
Thus, $v \approx 246$ GeV requires a dynamical mass
\be
m_{t \chi} \approx 0.6 \,{\rm TeV} 
\ee
for $M/m_{\chi \chi} \sim O(10)$ (values of $M/m_{\chi \chi}$ much 
larger than $O(10)$
would need fine-tuning in the gap equation for $m_{\chi \chi}$, while 
values below $\sim 5$ would need the inclusion of the next-to-leading 
order terms in the gap equations).
From eq.~(\ref{topmass}) it follows that a top mass of 175 GeV requires
\be
\frac{m_{\chi t}}{  m_{\chi \chi}} \approx 0.3 ~.
\label{ratio23}
\ee

One might worry at this stage that the four-fermion interactions (\ref{ops2}) 
and (\ref{ops3}), which are custodial--$SU(2)$
violating, inducing the top-bottom mass splitting,
may also lead to a large contribution to the electroweak 
parameter $T$ (equivalently, $\Delta\rho_*$). 
$T$ can be estimated in the fermion--bubble large--$N$ approximation as: 
\be
T \approx \frac{3 m_t^2}{16 \pi^2 \alpha(M_Z^2) v^2} 
\frac{m^2_{t \chi}}{m^2_{\chi t}} \left[
1 + O\left(m_{\chi t, \, t\chi}^2/m_{\chi \chi}^2\right) \right] ~,
\ee
where $\alpha$ is the fine structure constant.
The two-loop corrections to $T$ involving the operators 
(\ref{ops1}), which are significant for $M \sim 1$ TeV as discussed 
in ref.~\cite{cdt}, are small in our case due to the large $M$
(compared to the result in ref.~\cite{cdt}, there is a suppression of 
$(1 \, {\rm TeV}/M)^2 \lae 10^{-3}$, and
an enhancement of $(m_{t \chi}/m_t)^4 \sim 100$ from $\chi$ loops).
Moreover, the $S$ parameter is small, the leading contributions coming from 
a $\chi-t$ loop diagram, of order $(1/2\pi)(m_{\chi t}/m_{\chi\chi})$.
The fit to the electroweak data of the $S$ and $T$
parameters in the standard model with a Higgs mass of 300 GeV 
gives \cite{negs} a 1$\sigma$ ellipse in the $S-T$ plane whose projection on 
the $T$-axis is $T = 0.03 \pm 0.34$.
Requiring that our model does not exceed the 
1$\sigma$ upper bound on $T$, we obtain the following constraint:
\be
\frac{m_{t \chi}}{m_{\chi t}} \le 0.55 ~.
\label{ratio12}
\ee
The mass ratios (\ref{ratio23}) and (\ref{ratio12}) can arise naturally.
For example,
the set of values $m_{t \chi} \approx 0.5$ TeV, $m_{\chi t} \approx 0.9$ TeV,
$m_{\chi \chi} \approx 2$ TeV, $M \sim 50 - 70$ TeV would require 
$\kappa_1 \approx 0.02$,
$\kappa_{B-L} \approx 0.3$ and $1 - 8\kappa/(3\pi) \approx 0.02$, for $x = - 1/6$ and 
$\mu_{\chi\chi} \approx 30$ GeV. Note that the vectorlike quark has a mass 
approximately equal to $m_{\chi \chi}$, and therefore the phenomenological
constraints on its mass \cite{singlet} are satisfied. A consequence of the 
mixing between the $t$ and $\chi$ quark eigenstates is that the $V_{tb}$ 
element of the CKM matrix is given by the cosine of the mixing angle.
Therefore, the decrease in $|V_{tb}|$ compared to the standard model value 
is given by $m_{t \chi}^2/(2 m_{\chi \chi}^2) \sim 1\%$, which is potentially 
relevant for the single top production at the Run 3 of the Tevatron.

This model also implies the existence of pseudo-NGB's.
The chiral symmetry of the four-fermion interactions (\ref{ops1})-(\ref{ops3}),
is $SU(2)_W \times U(1)_{\chi_L} \times U(1)_{\chi_R} 
\times U(1)_{t_R}\times U(1)_{b_R}$, where the four $U(1)$'s are
the chiral symmetries of $\chi_L$, $\chi_R$, $t_R$ and $b_R$. One of the 
linear combinations of the $U(1)$ generators corresponds to the hypercharge 
gauge group $U(1)_Y$, while the remaining combinations are generators of
global $U(1)$'s.
The condensates $\langle \overline{\chi}_L \chi_R \rangle$
and $\langle \overline{\chi}_L t_R \rangle$ 
break $U(1)_{\chi_L} \times U(1)_{\chi_R} \times U(1)_{t_R}\times U(1)_{b_R}$ 
down to $U(1)_Y\times U(1)_{b_R}$ giving rise to two neutral NGB's which are 
linear combinations of $\overline{\chi} \gamma_5 \chi$, 
$\overline{\chi} \gamma_5 t$ and $\overline{t} \gamma_5 t$. 
These NGB's receive masses of order
\be
M_{\rm NGB} \approx \sqrt{\mu_{\chi A} m_{\chi A}} ~,
\ee
where $A$ stands for $\chi$ or $t$,
due to the explicit breaking of the global $U(1)$'s by the Dirac mass terms
in eq.~(\ref{lmass}). 
Note that the lower bound on the mass of a neutral 
pseudo-NGB that couples only to the top quark is much below the bound on the 
Higgs boson mass \cite{pgb}, so that even a current mass $\mu_{\chi t}$ 
of order 1 GeV is sufficient.

In summary, we have presented a simple mechanism of electroweak 
symmetry breaking 
based on the dynamics of the top quark with a seesaw
mechanism involving a vectorlike quark.
The model produces an acceptable top--quark mass,
is consistent with other electroweak data,
and does not require excessive fine-tuning.
It remains  to understand the necessary extension
of the scheme to give masses and mixing angles to all quarks and
leptons, and to construct attractive schemes for topcolor
breaking.  

{\it Acknowledgements:} We would like to thank Ken Lane and especially 
Sekhar Chivukula for useful comments on the manuscript.


\vfil
\end{document}